\documentclass[conference]{IEEEtran}
\IEEEoverridecommandlockouts
\usepackage{cite}
\usepackage{amsmath,amssymb,amsfonts}
\usepackage{algorithmic}

\usepackage{textcomp}
\usepackage{xcolor}
\def\BibTeX{{\rm B\kern-.05em{\sc i\kern-.025em b}\kern-.08em
    T\kern-.1667em\lower.7ex\hbox{E}\kern-.125emX}}
\usepackage{listings}
\usepackage{graphicx}
\usepackage{multirow}
\usepackage{amsmath}
\usepackage{bm}
\usepackage{comment}
\usepackage{array, makecell}
\usepackage{tabularx}
\usepackage[section]{placeins}
\usepackage[utf8]{inputenc}
\usepackage{hyperref}
\usepackage{xspace}
\usepackage{enumitem} 
\usepackage{todonotes} 
\usepackage{lipsum}
\usepackage{dblfloatfix}

\definecolor{mygreen}{HTML}{00CC00}

\newcommand{\code}[1]{\texttt{#1}}

\newcommand{\ie}{i.e.,~}
\newcommand{\wrt}{w.r.t.~}
\newcommand{\st}{s.t.,~}
\newcommand{\eg}{e.g.,~}

\newcommand{\Eg}{For example,~}
\newcommand{\cf}{cf.,~}
\newcommand{\etc}{etc.,~}
\newcommand{\qq}[1]{``#1''}
\newcommand{\Approach}[1]{$A_{#1}$}
\newcommand{\AV}{AV\xspace}
\newcommand{\QAnswer}{QAnswer\xspace}

\newcommand{\InformationRetrieval}{Information Retrieval\xspace}

\newcommand{\AnswerValidation}{Answer Validation\xspace}

\newcommand{\NLG}{NLG\xspace}
\newcommand{\GloVe}{GloVe\xspace}
\newcommand{\ELMo}{ELMo\xspace}
\newcommand{\wordtovec}{word2vec\xspace}

\newcommand{\Precision}{Precision\xspace}
\newcommand{\Recall}{Recall\xspace}
\newcommand{\Fscore}{F1 Score\xspace}
\newcommand{\KGQA}{KGQA\xspace}
\newcommand{\QA}{QA\xspace}

\newcommand{\ODQA}{ODQA\xspace}

\newcommand{\Wikidata}{Wikidata\xspace}

\newcommand{\KG}{KG\xspace}
\newcommand{\SPARQL}{SPARQL\xspace}
\newcommand{\Table}{Table\xspace}

\newcommand{\VANiLLa}{VANiLLa\xspace}
\newcommand{\dbo}[1]{\href{http://dbpedia.org/ontology/#1}{\texttt{dbo:#1}}}
\newcommand{\dbr}[1]{\href{http://dbpedia.org/resource/#1}{\texttt{dbr:#1}}}

\newcommand{\researchquestion}[1]{\textbf{RQ$_#1$}}

\begin{document}
\title{
Improving the Question Answering Quality using Answer Candidate Filtering based on Natural-Language Features
} 
\date{June 2021}

\author{\IEEEauthorblockN{1\textsuperscript{st} Aleksandr Gashkov$^\text{*}$}
\IEEEauthorblockA{\textit{Humanitarian Faculty} \\
\textit{Perm National Research Polytechnic University}\\
Perm, Russia \\
gashkov@dom.raid.ru}
\and
\IEEEauthorblockN{2\textsuperscript{nd} Aleksandr Perevalov$^\text{*}$}
\IEEEauthorblockA{\textit{Computer Science and Languages} \\
\textit{Anhalt University of Applied Sciences}\\
Köthen (Anhalt), Germany \\
aleksandr.perevalov@hs-anhalt.de}
\and
\IEEEauthorblockN{3\textsuperscript{rd} Maria Eltsova}
\IEEEauthorblockA{\textit{Humanitarian Faculty} \\
\textit{Perm National Research Polytechnic University}\\
Perm, Russia \\
maria\_eltsova@mail.ru}
\and
\IEEEauthorblockN{4\textsuperscript{th} Andreas Both}
\IEEEauthorblockA{\begin{tabular}{cc}\textit{Computer Science and Languages} & \textit{Technology Innovation Unit}\\
\textit{Anhalt University of Applied Sciences} &  \textit{DATEV eG} \\
Köthen (Anhalt), Germany & Nuremberg, Germany\\
andreas.both@hs-anhalt.de & andreas.both@datev.de
\end{tabular}
}
$^\text{*}$ corresponding authors
}
\maketitle

\begin{abstract}
Software with natural-language user interfaces has an ever-increasing importance.
However, the quality of the included Question Answering (\QA) functionality is still not sufficient regarding the number of questions that are answered correctly.

In our work, we address the research problem of how the \QA quality of a given system can be improved just by evaluating the natural-language input (i.e., the user's question) and output (i.e., the system's answer).

Our main contribution is an approach capable of identifying wrong answers provided by a \QA system.
Hence, filtering incorrect answers from a list of answer candidates is leading to a highly improved \QA quality. 
In particular, our approach has shown its potential while removing in many cases the majority of incorrect answers, which increases the \QA quality significantly in comparison to the non-filtered output of a system.
\end{abstract}


\section{Introduction}
 
\begin{figure*}[bt!]
    \centering
    \includegraphics[width=1\textwidth]{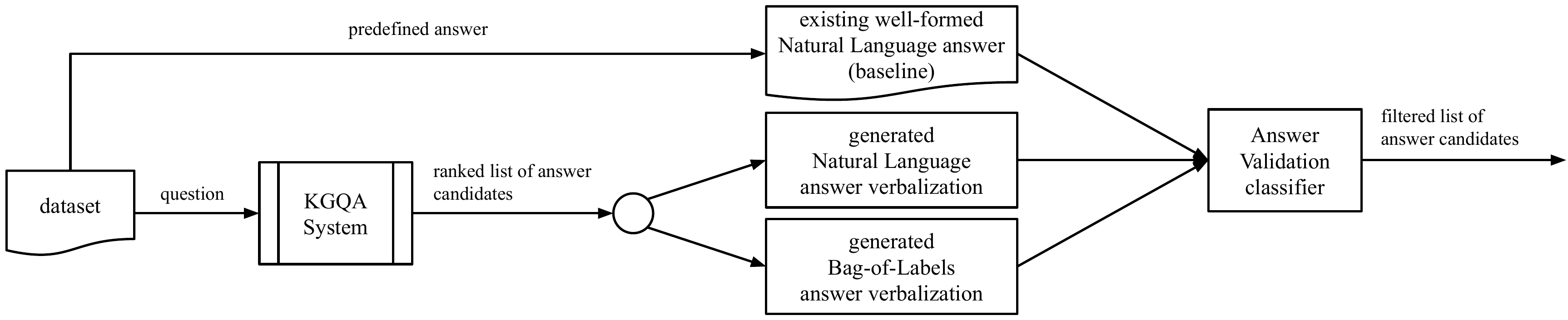}
    \caption{Overview of the general research idea}
    \label{fig:overview}
\end{figure*}
\begin{table*}[bt!]
\centering\renewcommand\cellalign{lc}
\setcellgapes{1.5pt}\makegapedcells
\centering
  \caption{\AnswerValidation publications of the last decade \cite{surveyOnAnswerValidation}}
  \label{tab:related_work}
  \resizebox{\textwidth}{!}{
  \begin{tabular}{|c|c|c|c|c|c|}
    \hline
    \textbf{Ref.} & \textbf{Year} & \textbf{Languages} & \textbf{Datasets} & \textbf{Methods} & \textbf{Evaluation score} \\
    \hline
    \cite{avQAIR} & 2010 & English, Spanish & ResPubliQA \cite{ResPubliQA} & EAT, NER, Acronym Checking & \makecell[c]{65\% Accuracy (English) \\ 57\% Accuracy (Spanish)} \\
    \hline
    \cite{TllezValero2010TowardsMQ} & 2010 & Spanish & CLEF 06 \cite{CLEF06} & RTE & 53\% Accuracy \\
    \hline
    \cite{DBLP:conf/clef/BabychHPP11} & 2011 & German & CLEF 11 \cite{CLEF11} & Rule-set & 44\% Accuracy \\
    \hline
    \cite{6040496} & 2011 & French & Web & Decision Tree Combination & 53\% MRR \\
    \hline 
    \cite{6406672} & 2012 & German & CLEF-QA \cite{CLEFQA} & LogAnswer Framework & 61\% Correct Top Rank \\
    \hline
    \cite{Pakray2012SemanticAV} & 2013 & English & AVE 08 \cite{AVE08} & RTE & \makecell[c]{58\% Precision \\ 22\% F-Score} \\
    \hline
    \cite{10.1007/978-3-642-38989-4_36} & 2013 & English & CLEF 11 \cite{CLEF11} & RTE & 45\% Precision \\
    \hline
    \cite{answerValidationForRussian} & 2013 & Russian & ROMIP \cite{ROMIP} & RTE & 70.4\% F-Score \\
    \hline 
    \cite{zamanov-etal-2015-voltron} & 2015 & English & Sem-Eval 2015 \cite{semEval15} & Word vectors + Classifier & \makecell[c]{62.35\% Accuracy \\ 46.07\% F-Score}\\ 
    \hline
    \cite{IKnowThereIsNowAnswer} & 2018 & English & SQuAD-T \cite{IKnowThereIsNowAnswer}  & \GloVe + GRU & 71.24\% F-score  \\ 
    \hline
     \cite{Hu2019ReadV} & 2019 & English & SQuAD 2.0 \cite{squad20}  & RTE + ELMo + Verifier & 74.2\% F-score\\ 
    \hline
\end{tabular}}
\label{tatriple to nlb:QA_multilingual_options}
\end{table*}

Question Answering (\QA) aims to provide precise answers to questions formulated in a natural language (NL). 
There are many different approaches for implementing \QA systems (\eg AskNow~\cite{AskNow}, QAnswer~\cite{QAnswer}, DrQA~\cite{DrQA}) focusing on specific paradigms and technologies. 
The two main directions of \QA are Open Domain Question Answering (\ODQA) and Knowledge Graph Question Answering (\KGQA). 

The common data flow in \QA systems is from a given input (question) to an output (answer). 
Typically, both are given using NL, \st the accessibility of such a system for humans is high.
In this work, we tackle the problem of answer validation (\AV) considering a \KGQA system as a black box. 
Thus, the concrete approach for computing the answer is hidden (\ie not visible and changeable) and only textual representation of question and answer are provided for further analyses. 

Here, we propose an approach for improving \QA quality by filtering out incorrect answer candidates.
The approach is built on the assumption that a considered \KGQA system provides a list of query candidates (\eg written in \SPARQL) that will be used to retrieve the answers from a knowledge graph (\KG).
In this scenario, the provided query candidates are executed and, therefore, it is decided whether they will retrieve the correct answer or not (\ie the incorrect answer candidates should be eliminated) -- we call this process answer validation (\AV) or filtering. 
Consequently, each query candidate must be transformed to an answer candidate in NL (\ie verbalized), in this regard, our \AV experiments are performed on 3 levels (\cf Figure~\ref{fig:overview}) where the verbalizations of query candidates (\ie NL form of a query) are
(\Approach{1}) provided using well-formed NL (written by a human as a baseline), 
(\Approach{2}) computed using Natural Language Generation (\NLG) considering the contained facts, and 
(\Approach{3}) computed using a bag-of-labels approach of available entities.

In this paper, we follow our long-term research agenda of improving the overall quality of \KGQA systems following a domain-agnostic approach that is not limited to just a single class of \KGQA systems.
Therefore, while having limited access to internal data structures of a \KGQA system, the NL form of the questions and answers gains importance. 
To show the significance of our approach, we not only consider \AV module quality (\ie \Fscore) but also its impact on the end-to-end \QA quality (\ie Precision@k, NDCG@k).



In this paper, we address the following research questions considering the task of filtering NL answer candidates:
\begin{enumerate}[label=\researchquestion{\arabic*}]
    \item Is it possible to improve the \QA quality while filtering answers just by their NL representation?
    \item What \QA quality is achievable while filtering the well-formed NL representations of the answer candidates (\cf Approach \Approach{1})?
    \item What \QA quality is achievable while having automatically generated NL answer candidates (\cf Approach \Approach{2} and Approach \Approach{3})?
\end{enumerate}


We conclude our main contributions as follows:
\begin{itemize}
    \item We demonstrate and evaluate two answer verbalization techniques (\Approach{2} and \Approach{3}) based on provided \SPARQL query candidates.
    \item We propose and validate a system-agnostic approach to filter a set of answer candidates using just its NL representation.
    \item The experimental results show that the proposed approach significantly improves the \QA quality.
\end{itemize}




This paper is structured as follows: 
after presenting the related work (Section~\ref{sec:relatedword}), we give an overview of our approach in Section~\ref{sec:approach}.
Section~\ref{sec:MaterialAndMethods} presents the used components and datasets.
Our experiments are described in Section~\ref{sec:exeriments} followed by their evaluation (Section~\ref{sec:analysis_discussion}). 
Section~\ref{sec:conclusion} concludes and outlines future work.
%
\section{Related Work}\label{sec:relatedword}
Techniques that tackle the task of validating the answer were applied mainly in \ODQA where systems are often required to rank huge amounts of candidate answers \cite{magnini-etal-2002-right}, \eg the incorrect answer candidates in form of textual paragraphs have to be eliminated by the \AV module. 
In our work, we also use a textual representation to validate answer candidates while using \KGQA systems. 
Thus, the \ODQA field is related to our research questions.

The comprehensive list of the \AV approaches was presented in \cite{surveyOnAnswerValidation}. 
Based on this we show the most related work in the \AnswerValidation field of the past decade in \Table~\ref{tab:related_work}.

In \cite{avQAIR} \AnswerValidation is applied to an \InformationRetrieval system. 
The validation process is performed on the basis of Expected Answer Type, Named Entities Presence, and Acronym Checking (only if a question is about an acronym). 
The authors mention that sometimes \AV module is \qq{too strict}, \ie it removes also correct answers. 

Other publications (\eg \cite{TllezValero2010TowardsMQ,Pakray2012SemanticAV,10.1007/978-3-642-38989-4_36,answerValidationForRussian}) are based on the idea of recognizing the textual entailment (RTE) using the output of several \QA systems, universal networking language, lexical similarity, and dependency parsing algorithms accordingly. 
All mentioned papers used textual datasets (containing a pair of one question and one answer).

Babych et al.~\cite{DBLP:conf/clef/BabychHPP11} utilized a rule-based decision algorithm that includes: syntactic analysis, predicate-argument relation analysis, and semantic relation analysis.
In \cite{6040496} Decision Tree Combinations are used that are based on the extracted textual features. 
Additionally, the system filters answer candidates (\ie potential answers to a given question) using Named Entities. 
The authors of \cite{6406672} applied the so-called LogAnswer Framework which incorporates case-based reasoning. 

\cite{zamanov-etal-2015-voltron,IKnowThereIsNowAnswer} and \cite{Hu2019ReadV} applied supervised learning methods. 
The first utilized the following features: lexical, n-grams, bad-answer specific, named entities, term-frequency vectors, and \wordtovec \cite{Word2Vec}.
In contrast, \cite{IKnowThereIsNowAnswer,Hu2019ReadV} used a deep neural network (DNN) model and only \GloVe~\cite{GloVe} and \ELMo~\cite{ELMo} features correspondingly.

To conclude we see how \AV approaches changed in time from rule-based to DNNs. 
Additionally, we recognize that \AnswerValidation was used mainly in \ODQA systems while -- to the best of our knowledge -- no work related to \KGQA exists.

\section{Approach}\label{sec:approach}



We assume here that a \KGQA system provides a ranked list of potential answers; we call them \emph{answer candidates}.
The first element of the list is reflecting the answer the \QA system would show to the user.
However, any answer in the list might be correct (\ie it is considered to correctly answering the given question), we denote such an answer candidate as \emph{correct answer}).
Note that there may be multiple candidate answers in the ranked list that are considered to represent the correct answer.
The other answer candidates are representing potential answers to the given question where the system has decided that they are not correctly answering the given question (\ie they are considered to be \emph{incorrect answers} \wrt the given question).


To evaluate our research idea, we utilize a text classification model trained on NL representations of correct and incorrect question-answer pairs, \ie the goal is to predict if a corresponding answer is correct for a given question using their NL representations (\cf~Figure~\ref{fig:overview}).

We use the following three settings to establish experiments where a verbalization of an answer is compared to the corresponding NL question: 
\Approach{1} a well-formed answer text, 
\Approach{2} an NL representation of each answer candidate, and
\Approach{3} a bag-of-labels representation of each answer candidate. 
These three approaches allow a distinct validation of the capabilities of our approach while validating it from a \qq{perfect} to a \qq{clumsy} textual answer representation.
The \AV classifier is trained \wrt each verbalization technique, and then the results are compared.
The \AV classifier is used to filter incorrect answer candidates, \st the \QA quality before and after the filtering process is measured.

For \Approach{1} a dataset is available (\cf Section~\ref{sec:MaterialAndMethods}).
For \Approach{2} and \Approach{3} we require an automatically executable process to generate the textual representations.


In the following, we will describe the different approaches.

\subsection{Approach 1 -- Well-formed NL Answers}
The first approach (\Approach{1}) should use well-formed NL answers to reflect a high-quality answer verbalization.
For example, a well-formed answer to the question \qq{What was the cause of death of John Kennedy?} can be \qq{John Kennedy was assassinated.}.

Therefore, it has high demands \wrt the textual answer quality and is hard to automate. 
For this reason, we will use the existing dataset \VANiLLa (\cf Section~\ref{sec:VANILLA}) containing well-formed NL answers to establish a baseline for our evaluations.

\subsection{Approach 2 -- Automatically generated NL answers}
The second approach (\Approach{2}) has less strict requirements rather than the first one to enable an automatic approach.
We will use an \NLG method to generate the answer verbalization.
We assume here that a \KGQA system (like \QAnswer, \cf Section \ref{sec:QAnswer}) provides a ranked list of potential answers in the form of \SPARQL queries (\ie a ranked list of \emph{query candidates}).
Hence, the \SPARQL queries are usable to generate the verbalizations with existing NLG tools (\eg Triple2NL, \cf Section~\ref{sec:Triple2NL}).
Regarding the example question (see above), the following \SPARQL can be proposed (see Listing~\ref{lst:approachSPARQL})\footnote{In this example, the \SPARQL query retrieves the correct answer regarding the given question. 
}.

\setcounter{lstlisting}{0}
\begin{lstlisting}[captionpos=b, caption=Retrieving answer URI via SPARQL from the DBpedia KG., label=lst:approachSPARQL]
# question: 
#     What was the cause of death of John Kennedy?
# query:
PREFIX dbr: <http://dbpedia.org/resource/>
PREFIX dbo: <http://dbpedia.org/ontology/>
SELECT ?answer WHERE { 
    dbr:John_F._Kennedy dbo:deathCause ?answer . 
}
# the result is dbr:Assassination_of_John_F._Kennedy
\end{lstlisting}

Providing the query and the result (RDF resource), we expect to receive the textual output from an \NLG tool. 
Given the example, a possible verbalization might be \qq{The John F. Kennedy's death cause is Assassination of John F. Kennedy}\footnote{The title of \dbr{dbr:Assassination\_of\_John\_F.\_Kennedy} is \qq{Assassination of John F. Kennedy}}.

\subsection{Approach 3 -- Answer verbalization based on bag-of-labels}

The third approach (\Approach{3}) manifests a number of labels to represent a simple verbalization of each answer candidate.
The labels are retrieved from the resources, properties, and results mentioned in a \SPARQL query candidate and its results (\cf the previous subsection).
The representation of such an answer is readable, however, it can barely be described as an NL sentence.

For example, considering the example in Listing~\ref{lst:approachSPARQL}, the following resources are present in the \SPARQL query and it's result: \dbr{John\_F.\_Kennedy}, \dbo{deathCause}, and \dbr{Assassination\_of\_John\_F.\_Kennedy}. 
After that, all the English labels of each resource are retrieved and concatenated (\ie a single string for each answer candidate).
In this example, the following verbalization of the answer candidate is obtained \qq{John F. Kennedy death cause Assassination of John F. Kennedy}.

\section{Materials and Methods}\label{sec:MaterialAndMethods}
In this section, we describe the components used to manifest the experimental environment. 
\subsection{The \VANiLLa Dataset}\label{sec:VANILLA}
Based on our preliminary research, it was decided to use the \VANiLLa dataset \cite{vanilla} that contains NL representation for both questions and answers 
in contrast to many other known datasets (\cf \cite{kacupaj2021paraqa}\footnote{The comparison \cite{kacupaj2021paraqa} reveals the lack of datasets with verbalized answers; however, the 
existing datasets (\eg VQuAnDa and ParaQA) with NL representation of an answer in the full form are relatively small: only 5000 question-answer pairs.}). 
It contains 107,166 examples with an 80\% (train) – 20\% (test) split of questions and answers in a full textual form. 
The dataset was built on top of the \Wikidata knowledge graph.
There are six fields for each instance of the \VANiLLa dataset: \texttt{question\_id}, \texttt{question}, \texttt{answer}, \texttt{answer\_sentence}, \texttt{question\_entity\_label}, \texttt{question\_relation}. 
Obviously, the dataset represents only the \emph{correct question-answer pairs}, which does not allow us to train the \AV classifier properly.
To generate the data for the 
\emph{incorrect question-answer pairs}
(\ie negative sampling), we created a set of the corresponding question-answer pairs by having randomly paired a question and an answer from different records of the \VANiLLa dataset. 
The method is precisely introduced in \cite{Both_2021}.
\subsection{The KGQA system \QAnswer}\label{sec:QAnswer}
Our study requires a \KGQA system to be utilized in order to prove the approach validity with actual (correct and incorrect) answer candidates.
Therefore, we selected the well-known \QAnswer system \cite{QAnswer} which represents the current state-of-the-art in \KGQA and satisfies the requirements for \Approach{2} and \Approach{3}.
\QAnswer provides an API to ask a question and receive the corresponding ranked query candidate list. 
It supports questions over several knowledge bases including \Wikidata. 
The \SPARQL queries are constructed by traversing the KG and discovering how the concepts and relations that are mentioned in the question are arranged. 
To the best of our knowledge, there is no other \KGQA system available that satisfies our requirements and provides a state-of-the-art \QA quality.
\cite{korablinov2020rubq} compares the results of DeepPavlov’s and QAnswer’s top-1 results on RuBQ dataset where QAnswer outperforms DeepPavlov in terms of most metrics.
\subsection{Triple-based NLG using Triple2NL}\label{sec:Triple2NL}
While following our approach \Approach{2}, we require automatically generated NL representations for query candidates.
However, there are quite a few state-of-the-art open-source toolkits that introduce an API or a module for programming language. 

As the used \QA system (\QAnswer, \cf the previous subsection) responds with a list of query candidates, a tool matching most of our demands is Triple2NL.
It represents an integrated system\footnote{It provides a REST API, which can be deployed locally, but it is not able to generate text from more than one triple.} generating a complete NL representation for RDF and SPARQL.
The results (\cf \cite{ngomo2019holistic,ferreira20202020}) demonstrate that Triple2NL generates comprehensible texts.
Triple2NL's approach is based on a bottom-up process to verbalization and exploits some rules.
To create a textual representation, Triple2NL tries to get the label or the name of all triple elements from DBpedia. 
If no label nor name is found, the last part of URI is taken as an NL representation. This allows us to generate text for any \KG by submitting labels instead of URIs.
\subsection{BERT Model as Answer Validation Classifier}\label{sec:textClassification}



Modern text classification methods utilize large sets of unstructured data to pre-train a model. 
In recent years, transformer-based models are holding a significant part in the whole NLP industry and research community. 
Such models as ELMo~\cite{ELMo}, ULM-FIT~\cite{ULM-FIT}, XLNet~\cite{XLNet}, and BERT~\cite{BERT} represent the current state-of-the-art. 
The well-known BERT still stays as the landmark model showing one of the best results in many downstream tasks (\cf~\cite{HowToFineTuneBERT}).

In this work, we used the \texttt{bert-base-cased} model\footnote{Available online at \url{https://huggingface.co/bert-base-cased}.} the next sentence prediction (NSP) setting as the input consists of a textual tuple question-answer.
\section{Experimental Setup}\label{sec:exeriments}
\subsection{Answer Validation Classifier}
The \AV classifier model (see Section \ref{sec:textClassification}) was trained and evaluated according to each answer verbalization approach separately (see Section \ref{sec:approach}).
Each experiment was executed ten times using different random seed values for the dataset's shuffle to ensure validity.
We used such well-known metrics for classification as \Precision, \Recall, and \Fscore.
Given the obtained metric results, we calculated standard deviation (std) which is denoted as \qq{$\pm$} in our results.

The generated dataset consists of the mixed correct and incorrect subsets and contains 66,632 records from the original training subset of \VANiLLa (the original test subset was used in the next experiment described in Section \ref{ssec:experiment_influence_of_av_on_kgqa}).
The train/test split was done randomly in every experiment with the corresponding ratio 67\%/33\%.
We evaluated both balanced and unbalanced (50 incorrect to 1 correct tuple) dataset's settings to see the corresponding effect.


\subsection{Evaluation of Answer Validation's impact on Question Answering Quality}\label{ssec:experiment_influence_of_av_on_kgqa}

\begin{figure*}[tb!]
    \centering
    \includegraphics[width=\textwidth]{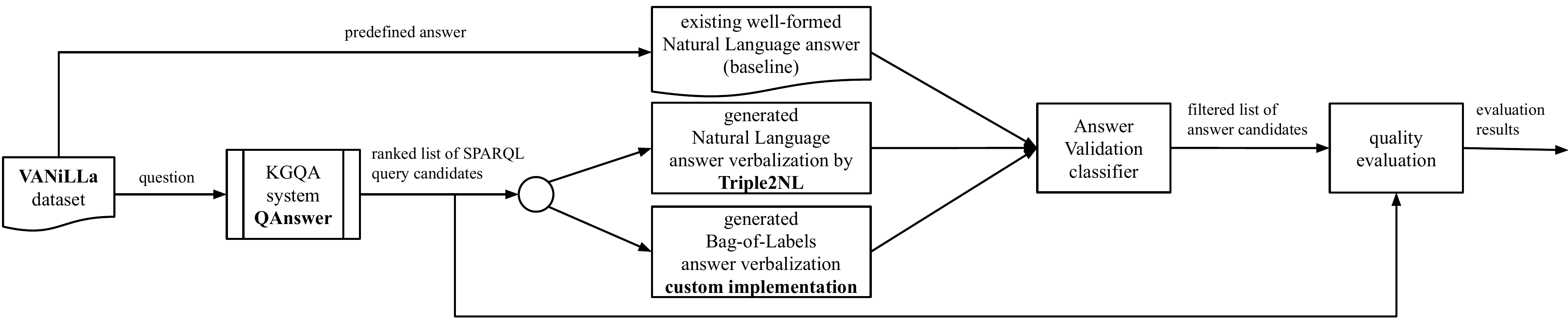}
    \caption{Flowchart of the \QA quality comparison process. The quality is compared between filtered and non-filtered answer candidate sets}
    \label{fig:filteringAnswersQA}
\end{figure*}

To demonstrate the impact of \AV on a \KGQA system's quality, we used the following approach 
(see Figure~\ref{fig:filteringAnswersQA}). 
The \KGQA system was evaluated on the original test subset of \VANiLLa containing 21,360 records, but only 8,955 of them have unique questions. 
We decided to send each question to \KGQA only once as the response is stable (\ie not changing inside one session).
A considered \KGQA system (here: \QAnswer) processed the test dataset and the obtained ordered lists of query candidates were cached.
As the query candidate was presented in a form of \SPARQL, we were able to apply \Approach{2} and \Approach{3} to validate the candidates.
Using one of the selected verbalization approaches, the answer candidates were being filtered by the \AV classifier while the order of candidates stayed the same (\ie we removed such answer candidates from the list).
Finally, the \QA quality was measured and compared based on the non-filtered and filtered answer candidate sets.
We used well-known metrics such as Precision@k (P@k), Normalized Discounted Cumulative Gain@k (NDCG@k) with $k \in \{1, 5\}$ as \QA quality measures \cite{schutze2008introduction}.


\section{Results and Discussion}\label{sec:analysis_discussion}
\subsection{Experiment 1: High quality NL questions and answers (\Approach{1})}
The first experiment is intended to evaluate the capabilities of the \AV component on well-formed answer representations which are retrieved from the \VANiLLa dataset.
The results of the experiment are presented in Table~\ref{tab:Experiment1}.


\begin{table}[!h]
\centering
  \caption{Detailed results for Experiment 1}
  \label{tab:Experiment1}
  \begin{tabular}{|c|c|c|}
    \hline
    \textbf{F1 Score} & \textbf{Precision} & \textbf{Recall} \\
    \hline
    \hline
    \multicolumn{3}{|c|}{balanced data (1 to 1)} \\
    \hline
    \textbf{0.9968} $\pm$ 0.0089 & \textbf{0.9968} $\pm$ 0.0100 & \textbf{0.9968} $\pm$ 0.0098 \\
    \hline
    \multicolumn{3}{|c|}{unbalanced data (50 to 1)} \\
    \hline
    0.9838 $\pm$ 0.0276 & 0.9918 $\pm$ 0.0251 & 0.9761 $\pm$ 0.0454 \\
  \hline
\end{tabular}
\end{table}

The first experiment's results are strong and fulfill the quality requirements.
They demonstrate that detecting incorrect question-answer pairs using only text is possible, and high quality is achievable. 
Although, using the unbalanced setting for training negatively affects mostly the recall which is the most important metric for the intended goal.
Hence, assuming NL answers of very high quality are provided by a QA system, then our approach should be capable of identifying (and therefore, filtering) incorrect answers.




\subsection{Experiment 2: NLG of limited quality}
In this setting, we are evaluating artificially generated NL answers.
There are computed automatically from \SPARQL queries and the corresponding results.


To our best knowledge, there is no \KGQA system available that is providing an API to produce full-fledged NL answers.
Consequently, it is required to generate answer verbalization from the available information (\ie \SPARQL query candidates).

Generating artificial answers in a three-step process includes (1) providing a question in textual form to the \KGQA system, (2) sending the computed \SPARQL query answer candidates to \Wikidata, and (3) generating NL representation from the obtained list of the query candidates and \Wikidata response.

The first step is straightforward: the question provided to the \KGQA system. 
The system returns a set of query candidates.
A typical \SPARQL query consists of one or more triples including at least one variable (\cf the following example). 
\begin{lstlisting}[captionpos=b, caption=An example SPARQL query over Wikidata., label=lst:wikidataSPARQLExample], 
PREFIX wd: <http://www.wikidata.org/entity/>
PREFIX wdt: <http://www.wikidata.org/prop/direct/>
SELECT DISTINCT ?o2 WHERE {
    ?s1  ?p1  wd:Q57747377 .
    ?s1  wdt:P21 ?o2 .
}  LIMIT 1000
\end{lstlisting}

The second step extracts all variables' values from the query's result set as all triples' positions must be filled to generate a text. 
Hence, we change \texttt{SELECT DISTINCT ?o2 WHERE} to \texttt{SELECT DISTINCT * WHERE} and execute a query on \Wikidata. 
For the given example it returns:
\begin{lstlisting}[captionpos=b, caption=Result set of the SPARQL query of Listing~\ref{lst:wikidataSPARQLExample} with \texttt{SELECT DISTINCT * WHERE}. The RDF vocabulary prefixes are ommited., label=lst:resultWikidataSPARQLExample]
[
 {
  "s1": "wd:Q16027703",
  "p1": "wdt:P735",
  "o2": "wd:Q6581097"
 },
 {    
  "s1": "wd:Q2976815",
  "p1": "wdt:P735",
  "o2": "wd:Q6581097"
 }
]
\end{lstlisting}

The third step is generating the NL representation. The SPARQL request is split into triples and all variables are substituted with their values. Each triple is converted into a text individually and then, if there is more than one triple, joined with the word \qq{and}. 
\Eg the first set of values gives a sentence: \qq{Claude-Nicolas Le Cat is given name Claude-Nicolas and Claude-Nicolas Le Cat's sex or gender is male.}
Although the sentence looks not fluent, it serves our goal well, which is clear from experimental results.

The experiment was carried out on balanced and unbalanced datasets.
The results are shown in Table~\ref{tab:Experiment2}.

\begin{table}[!h]
\centering
  \caption{Detailed results for Experiment 2}
  \label{tab:Experiment2}
  \begin{tabular}{|c|c|c|}
    \hline
    \textbf{\Fscore} & \textbf{Precision} & \textbf{Recall} \\
    \hline
    \hline
    \multicolumn{3}{|c|}{balanced data (1 to 1)} \\
    \hline
     \textbf{0.9982} $\pm$ 0.0011 & \textbf{0.9980} $\pm$ 0.0020 & \textbf{0.9983} $\pm$ 0.0008 \\
    \hline
    \multicolumn{3}{|c|}{unbalanced data (50 to 1)} \\
    \hline
     0.9631 $\pm$ 0.0170 & 0.9321 $\pm$ 0.0317 & 0.9968 $\pm$ 0.0053 \\
  \hline
\end{tabular}
\end{table}

The obtained results demonstrate that the difference between the well-formed textual answers and the generated answers quality metrics is not significant. 
However, in this experiment, the precision drops significantly for the unbalanced set.

\subsection{Experiment 3: Bag-of-labels representation}
In this experimental setting, we evaluate bag-of-labels answer verbalizations computed from entities and relations that are available in \SPARQL query and its result set.
The evaluation results of the \AV classifier according to \Approach{3} are shown in Table~\ref{tab:Experiment3}.
\begin{table}[!h]
\centering
  \caption{Detailed results for Experiment 3}
  \label{tab:Experiment3}
  \begin{tabular}{|c|c|c|}
    \hline
    \textbf{\Fscore} & \textbf{Precision} & \textbf{Recall} \\
    \hline
    \hline
    \multicolumn{3}{|c|}{balanced data (1 to 1)} \\
    \hline
    \textbf{0.9613} $\pm$ 0.0029 & \textbf{0.9355} $\pm$ 0.0109 & \textbf{0.9886} $\pm$ 0.0089 \\
    \hline
    \multicolumn{3}{|c|}{unbalanced data (50 to 1)} \\
    \hline
    0.9205 $\pm$ 0.0570 & 0.9289 $\pm$ 0.0520 & 0.9170 $\pm$ 0.0870 \\
  \hline
\end{tabular}
\end{table}

In comparison to the balanced dataset setting, usage of unbalanced data (50 incorrect to 1 correct) leads to a quality drop.
Additionally, the classification results demonstrate less robustness of the model while paying attention to standard deviation (as in the previous experiment).
These results show that it is still possible to filter incorrect answers just by having the entity labels of the answer candidate set.

\subsection{Filtering Answers for a Question Answering Process}

In this subsection, the impact of \AV on \QA quality is demonstrated.
We conducted the corresponding experiments using \Approach{2} and \Approach{3}, the results are shown in Table~\ref{tab:FilteringAnswersQA}.


\begin{table*}[!t]
\centering
  \caption{Comparison of \QA quality before and after filtering by our \AV classifiers}
  \label{tab:FilteringAnswersQA}
\begin{tabular}{c|c|c|c|c|c}
\multicolumn{2}{c|}{\textbf{P@1 = NDCG@1}}      &
\multicolumn{2}{c|}{\textbf{P@5}}               & \multicolumn{2}{c}{\textbf{NDCG@5}}            \\
\hline
\textbf{Before AV} & \textbf{After AV} & \textbf{Before AV} & \textbf{After AV} & \textbf{Before AV} & \textbf{After AV} \\
\hline
\multirow{4}{*}{0.2476} & \Approach{2} & \multirow{4}{*}{0.1036} & \Approach{2} & \multirow{4}{*}{0.3249} & \Approach{2} \\
\cline{2-2} \cline{4-4} \cline{6-6}
& \textbf{0.4251} & & \textbf{0.1368} &  & \textbf{0.4698} \\
\cline{2-2} \cline{4-4} \cline{6-6}
& \Approach{3} & & \Approach{3} & & \Approach{3} \\
\cline{2-2} \cline{4-4} \cline{6-6}
& 0.2948 & & 0.1183 & & 0.3787              
\end{tabular}
\end{table*}

The results provide strong evidence that the \QA quality was significantly increased while using both approaches in comparison to not-validated answer candidate sets.
While considering P@1 (NDCG@1), the results were improved from 0.2476 to 0.4251 (\ie by 71.7\%) while using \Approach{2} and to 0.2948 (\ie by 19.1\%) while using \Approach{3}.
The corresponding results \wrt the P@5 are as follows: from 0.1036 to 0.1368 (\ie by 32.0\%) while using \Approach{2} and to 0.1183 (\ie by 14.2\%) while using \Approach{3}.
Finally, the results for NDCG@5 are: from 0.3249 to 0.4698 (\ie by 44.6\%) while using \Approach{2} and to 0.3787 (\ie by 16.5\%) while using \Approach{3}.

The \Approach{2} outperforms \Approach{3} which 
intuitively comes from the fact that the \Approach{2} produces more fluent answer verbalization rather than \Approach{3}.
Hence, the \AV classifier is able to capture more semantics and therefore, distinguish between correct and incorrect question-answer tuples.
The \Approach{1} obviously could not be evaluated as it would require a huge amount of manual work or a fine-tuned \NLG module.
However, we assume that it would produce even higher quality.

It is worth emphasizing that the \AV filtering removed the majority of the answer candidates.
For example, in the case of \Approach{2} 57 out of 60 candidates were removed on average, while in the case of \Approach{3} -- 44 out of 60.
Consequently, the position of correct candidates in the list goes up because of filtering out the incorrect candidates.

\subsection{Limitations}
Notwithstanding the promising results described in the previous subsection, our work has several limitations.
The generalizability of the study is limited by the used dataset (\VANiLLa), \KGQA system (\QAnswer), \NLG tool (Triple2NL), and model for \AV classifier (BERT).
The source data for \Approach{2} (\SPARQL queries) was constrained by removing \code{COUNT}, \code{MAX}, \code{LIMIT}, \code{ORDER} and \code{FILTER} clauses as Triple2NL can only verbalize simple RDF sequences.

In the subject-predicate-object structure, \VANiLLa defines resource URIs solely for the predicate, in contrast, only labels are given for the subjects and the objects of the sentences. 
Hence, it becomes difficult to determine whether an answer candidate is correct or not without any ambiguity \wrt to the uniqueness of labels in \Wikidata.
For example, there are 6172 resources with the label \qq{Correction} from the \VANiLLa test subset.
While it is hard to distinguish automatically entities with the same label for a given question, we considered any entity with a matching label as relevant for the current question.

There is another limitation in the \VANiLLa dataset's structure.
Some questions imply many correct answers, but only one of them is given as a reference in the dataset.
In this case, we believe that if an object's (or subject's) label and predicate correspond to the \VANiLLa record, then the answer is correct, regardless of the number of entities.

Obviously, these compromises might influence the quality (\ie there is an opportunity to improve our approach).



\subsection{Summary}

Despite several limitations described in the previous subsection, the answer validation approach demonstrated its effectiveness by eliminating incorrect answer candidates based only on NL representations of questions and possible answers.

The \AV classifier on well-defined answer verbalizations (\Approach{1}) as well as on the automatically generated answer verbalizations (\Approach{2}, \Approach{3}) demonstrated a reasonable \Fscore~-- 0.9968, 0.9982, and 0.9613 respectively for balanced data setup.
The unbalanced data setup leads to the significant decrease in the classification quality.
Therefore, the \AV classifier should be trained on a balanced dataset to avoid a high false-negative rate while filtering answer candidates.

Given the comparison results before and after answer filtering with \AV classifier, we were enabled to significantly improve the \KGQA system's quality.
The maximal improvement was achieved using \Approach{2}: +71.7\% (P@1/NDCG@1), +32.0\% (P@5), and +44.6\% (NDCG@5).
Although, \Approach{3} also demonstrated reasonable quality improvement.

While answering \researchquestion{1}, the \AV classifier is capable of eliminating wrong answers and, therefore, moving correct answers to the top of the result list.
The \AV classifier is capable of distinguishing between correct and incorrect answers based just on automatically generated verbalizations obtained within \Approach{2} and \Approach{3} (\researchquestion{3}).
We could not fully answer \researchquestion{2} as the manual NL generation for each query candidate requires huge labor costs.
Therefore, extrapolating classification results of \Approach{1} (Table~\ref{tab:Experiment1}) on \QA quality comparison experiments (Table~\ref{tab:FilteringAnswersQA}), we assume that the possible outcome of \Approach{1} may even outperform \Approach{2} and \Approach{3}.

\section{Conclusion}\label{sec:conclusion}
In this paper, we presented an approach to filter computed NL answer candidates within a \KGQA process.
We have purposely limited ourselves to use only NL features, in particular, labels of the entities, properties, and concepts, \st our approach is not tightly integrated into existing \KGQA systems.
Hence, our approach is a \emph{generalized method for optimizing \KGQA systems}.
For proving this, we defined 3 research questions that were answered by our experiments.
As we have used the state-of-the-art \KGQA system \QAnswer, we consider our results to be very solid, as we were capable of improving the quality even for this well-established system.
Hence, we can conclude that our approach should be applicable to any \QA system that provides additional information about the answer candidates.

Our results show a huge impact \wrt the \VANiLLa dataset.
In the executed experiments, we were capable of removing incorrect answer candidates (from the ranked list of SPARQL queries), \st the result improved by 71.7\% and 44.6\% (\wrt P@1 and NDCG@5).


In the future, we will integrate our approach into existing \QA systems that are using NL input and output, \st an additional QA quality improvement is achieved.
It would also be possible to use the results of our classification as an additional feature for the core of a \QA engine, \st the quality of answer candidates might be improved on a lower system level.
In particular, there are \QA systems that provide high performance while at the same time not interpreting the question's semantics precisely.
Such systems might benefit heavily from our approach.

\bibliographystyle{IEEEtran}
\bibliography{IEEEabrv,conference_101719}
\end{document}